\documentclass{edp-conf}
\usepackage{graphicx}
\begin{document}
\def\la{\buildrel<\over\sim}
\def\ga{\buildrel>\over\sim}

\TitreGlobal{SF2A 2005}

\title{ZAMS ROTATIONAL VELOCITIES OF Be STARS}

\author{Martayan, C.}\address{Observatoire de Paris, Section d'Astrophysique 
de Meudon, GEPI, FRE K 2459}
\author{Zorec, J.}\address{Institut d'Astrophysique de Paris, UMR7095 CNRS, 
Universit\'e Pierre \& Marie Curie}
\author{Fr\'emat, Y.}\address{Royal Observatory of Belgium}
\author{Hubert, A.M.$^1$}
\author{Floquet, M.$^1$}

\runningtitle{ZAMS rotation of Be stars}

\index{Martayan, C.}
\index{Zorec, J.}
\index{Fr\'emat, Y.}
\index{Hubert, A.M.}
\index{Floquet, M.}

\maketitle

\begin{abstract} We show that Be stars belong to a high velocity tail of a 
single B-type star rotational velocity distribution in the main sequence (MS). We 
studied 127 galactic field Be stars and obtained their true equatorial velocity
at the ZAMS using models of stellar evolution with rotation. There is a sharp 
mass-dependent cut in the ZAMS under which there is no Be star. Velocities 
above this cut follow a Gaussian-tail distribution. B stars with ZAMS 
rotational velocities lower than the cut probably cannot become Be.
\end{abstract}

\section{Motivation and Method}

 There is a long-lasting debate whether Be stars belong to a high velocity end
of a single distribution suited to all MS B stars, or they form a stellar 
group with a separate high-velocity distribution. To obtain answer elements to 
this question, we corrected the $V\!\sin i$ for gravitational darkening 
effects (Fr\'emat et al. 2005), calculated the inclination angle $i$ and 
derived masses and ages of 127 program Be stars using models of stellar 
evolution with rotation (Meynet \& Maeder 2000, Zorec et al. 2005). To derive
the respective ZAMS true equatorial velocities we have taken into account four
main phenomena determining the variation of the surface rotation with age 
(Meynet \& Maeder 2000): 1) the equatorial radius-dependent changes; 2) 
variations due to angular momentum-loss through mass-loss events; 3) 
variations carried by internal density distribution changes with age and 
rotation, which affect the stellar inertial momentum; 4) meridional 
circulation and other hydrodynamical instabilities that produce internal 
angular momentum redistribution with time scales of the order of $\tau_{\rm 
HK}/\eta$, where $\tau_{\rm HK}$ is the Kelvin-Helmholtz time and $\eta\sim$ 
0.9 is the ratio of centrifugal to gravitational acceleration.\par
 We note that $V_{\rm ZAMS}$ is the equatorial velocity the star acquires 
after its rapid initial short-lasting re-arrangement of the internal angular 
velocity law $\Omega(r)$. This re-arrangements last no more than 1 to 2\%
of the stellar MS life (Denissenkov et al. 1999, Meynet \& Maeder 2000) and 
transforms an initial entirely rigid rotation into another differential one 
where the stellar core has a faster angular velocity than the equator.

\section{Results and Conclusion}

 The transformation of individual $V\!\sin i$ parameters into $V_{\rm ZAMS}$ 
is shown in Fig. 1a. The most striking feature in this figure is the neat 
cut depicted by the regression line obtained from the $V_{\rm min}$ in each 
mass-interval. According to this finding, no Be star is seen below the cutting 
line. This indicates that stars need an initial velocity $V_{\rm ZAMS}\ga$ 
$V_{\rm min}$ in order to become Be at any moment in the MS evolutionary phase. 
In each mass-interval we divided the $V_{\rm ZAMS}$ by the corresponding
$V_{\rm min}$ and formed the global histogram shown in Fig. 1b. The fit that 
better describes the distribution obtained is a Gaussian tail. We note that
under the histogram there must be roughly 17\% of stars out of the whole MS
B-type star population. More than 80\% of B-type stars must then be gathered 
into the $V_{\rm ZAMS}/V_{\rm min}$ $\la 1$ interval. It may then happen that
Be stars do not form a separate distribution, but possibly a Gaussian-like tail 
of a more general distribution of rotational velocities that encompasses the 
whole MS B-type star population.\par
 We note that some B stars without emission lines, of which a non negligible 
proportion is in the lower mass-extreme ($M<7M_{\odot}$) is represented by Bn
stars, which distribute over both velocity intervals. Bn stars with $V_{\rm 
ZAMS}<V_{\rm min}$ might then never become Be, while those with $V_{\rm 
ZAMS}>V_{\rm min}$ could probably do.\par  

\begin{figure}[h]
\centering
\includegraphics[height=5cm,width=11cm]{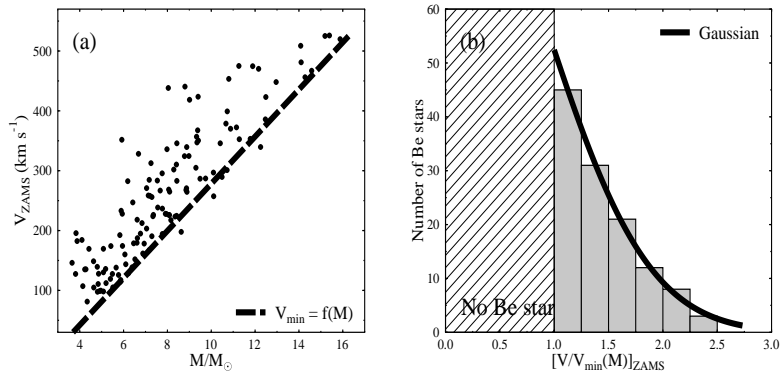}
\caption{(a): Distribution of true rotational velocities at ZAMS of the 
studied galactic Be stars. There is a clear cut under which there is no Be 
star. (b): Frequency distribution of $V_{\rm ZAMS}/V_{\rm min}$ ratios and
fit with a Gaussian distribution tail.}
\end{figure}


\begin{thebibliography}{}
\bibitem{}Denissenkov, P.A., Ivanova, N.S., \& Weiss, A. 1999, A\&A, 341, 181
\bibitem{}Fr\'emat, Y., Zorec, J., Hubert, A.M., \& Floquet, M. 2005, A\&A, 
440, 305
\bibitem{}Meynet, G., \& Maeder, A. 2000, A\&A, 361, 101
\bibitem{}Zorec J., Fr\'emat, Y., Cidale, L.  2005, A\&A, in press and 
astro-ph/0509119
\end{thebibliography}
\end{document}